# Temperature effect on the coupling between coherent longitudinal phonons and plasmons in *n*- and *p*-type GaAs

Jianbo Hu,[1,2,*] Hang Zhang,[1,2] Yi Sun,[1,2] Oleg V. Misochko,[3,4] Kazutaka G. Nakamura[5]

[1]State Key Laboratory for Environment-Friendly Energy Materials, Southwest University of Science and Technology, Mianyang, Sichuan 621010, China

[2]Laboratory for Shock Wave and Detonation Physics Research, Institute of Fluid Physics, China Academy of Engineering Physics, Mianyang, Sichuan 621900, China

[3]Institute of Solid State Physics, Russian Academy of Sciences, 142432 Chernogolovka, Moscow region, Russia

[4]Moscow Institute of Physics and Technology (State University), 141700 Dolgoprudny, Moscow Region, Russia

[5]Materials and Structures Laboratory, Tokyo Institute of Technology R3-10, 4259 Nagatsuta, Yokohama 226-8503, Japan

*To whom correspondence should be addressed. Email: jianbo.hu@caep.cn

**Abstract**

The coupling between longitudinal optical (LO) phonons and plasmons plays a fundamental role in determining the performance of doped semiconductor devices. In this work, we report a comparative investigation into the dependence of the coupling on temperature and doping in n- and p-type GaAs by using ultrafast optical phonon spectroscopy. A suppression of coherent oscillations has been observed in p-type GaAs at lower temperature, strikingly different from n-type GaAs and other materials in which coherent oscillations are strongly enhanced by cooling. We attribute this unexpected observation to a cooling-induced elongation of the depth of the depletion layer which effectively increases the screening time of surface field due to a slow diffusion of photoexcited carriers in p-type GaAs. Such an increase breaks the requirement for the generation of coherent LO phonons and, in turn, LO phonon-plasmon coupled modes because of their delayed formation in time.

**I. Introduction**

The interactions among elementary excitations in materials playing a fundamental role to determine physical properties remain the main concern of solid state physics. As one of them, the coupling between longitudinal optical (LO) phonons and plasmons in doped semiconductors, important for the performance of semiconductor devices, has been extensively studied in the last decades using either Raman spectroscopy or ultrafast coherent phonon spectroscopy [1-10]. Experiments demonstrate that the coupling most effectively occurs, when plasmons and LO phonons have comparable frequencies, to form LO phonon-plasmon coupled (LOPC) modes, which can be well described by a pair of time-independent coupled equations for the electronic polarization and lattice displacement [2,11-13]. Recent ultrafast time-domain experiments, however, show unambiguously that the coupling is a kinetic process, that is, the formation of coupled modes is not instantaneous but takes almost one period of collective excitations [14,15]. The delayed formation not only has a physical meaning in terms of quantum kinetics [16,17] but also affects remarkably coherent control of LOPC modes [10,18].

In spite of the significance of the coupling kinetics, little efforts have been paid to further explore its intrinsic features, for example, as a function of temperature or doping type, although a few frequency-domain results are available in the literature providing no kinetic information [4,6,19]. It is worth noting that Huber *et al.* [15] have shown the coupling kinetics to be sensitive to the photoexcited carrier density. However, it is still not well understood how the intrinsic doping influences the coupling and what is the exact temperature dependence.

In this work, we report on the ultrafast time-resolved investigation on coherent oscillations in n-type and p-type GaAs at various temperatures. The temperature dependence of these two specimens exhibits distinctly different behaviors. For n-type GaAs, coherent oscillations of LO phonons are enhanced with decreasing temperature while those of LOPC modes are only slightly modulated. In contrast, coherent oscillations of both modes in p-type GaAs are strongly suppressed at a lower temperature. We ascribe such an abnormal behavior to the increased duration of driving force, which degrades the generation of coherent phonons and, in turn, the plasmon-LO phonon coupling.

**II. Experimental method**

Ultrafast coherent phonon spectroscopy has been applied to probe the transient reflectivity change in a Zn-doped p-type ($N_p \geq 5\times10^{18}$ /cm$^3$) and a Si-doped n-type ($N_n = (1\pm0.5)\times10^{18}$ /cm$^3$) GaAs(100) (MTI Co.). The experimental details have been well described in Ref. [10]. Briefly, ultrashort laser pulses with the duration of 45 fs and the central wavelength of 800 nm, outputted by a Ti:sapphire oscillator at the repetition rate of 86 MHz, are divided into two to excite and probe the dynamics. The polarizations of pump and probe pulses are set orthogonal to each other to eliminate the possible coherent artifact. An electro-optic sampling scheme is adopted to detect the normalized pump-induced difference, $\Delta R_{eo}/R_0$, between the vertically and horizontally polarized components of the reflected probe light using a balanced photodetector, thus eliminating the intense isotropic change in the complex index of refraction [20]. Here, $R_0$ is the reflectivity without pump. In such a detection configuration, the signal change of $10^{-7}$ is detectable with a fast scan technique. The temperature is controlled in the range from room to liquid nitrogen temperature by attaching the sample to the cold finger of a cryostat (ColdEdge, Co.) with the accuracy better than 1 K. The fluences of the pump and probe pulses are 95 μm/cm$^2$ and 6 μm/cm$^2$, respectively, inducing only a weak perturbation of optical properties of the samples.

**III. Results and discussion**

The transient reflectivity changes induced by the pump pulse are shown in Fig. 1(a) and Fig. 1(b) (red solid lines) for n-type and p-type GaAs, respectively. In both samples, the changes are composed of oscillatory signals superimposed on a non-oscillatory background resulting from the interband excitation of valence electrons. Within the concerned temperature range, the photon energy (1.55 eV) of the pump pulse remains above the GaAs bandgap (1.42-1.51 eV) [21], thus creating electron-hole free carriers. The oscillatory change of the reflectivity carries the information of collective motions of lattices and charged particles. In n-type GaAs (Fig. 1(a)), the oscillatory signals show obvious beating pattern, indicating that more than one coherent mode has been excited. Fast Fourier transforms (FFT) of the time-domain signals, as shown in Fig. 1(c), exhibit exactly two modes at the frequency of 8.7 THz and 7.7 THz, corresponding to LO phonons and the lower branch of LOPC modes, respectively. This assignment was made according not only to the frequency and symmetry of each mode but also to the previous pump fluence dependent measurement [10]. Here, the frequency of LOPC modes is very close to the estimated one, 7.9 THz, from the total density, $(4.2\pm0.5)\times10^{18}$ /cm$^3$, of photoexcited and intrinsic carriers [22]. The generation mechanism of coherent phonons in GaAs has been attributed to transient screening of surface field [20]. With the decreasing temperature, coherent oscillations last longer (see Fig. 1(a)), showing an enhanced and narrowed FFT peak in the frequency domain (see Fig. 1(c)). Such a dependence has been often observed in other materials [23-28] and can be easily understood as a reduced anharmonic decay of optical phonons into acoustic phonons at the lower temperature. Regarding LOPC mode, its FFT peak has nearly been modulated by temperature.

In p-type GaAs at room temperature (Fig. 1(b)), the time-domain signal shows no beating. Instead, it starts with a fast damping (within 1 ps) oscillation, followed by a longer-lived (up to 7 ps) one. As shown in Fig. 1(d), its FFT exhibits two peaks, one of which is located at the frequency of LO phonons and the other is assigned to LOPC mode broadened by strong damping [10]. With the decreasing temperature, coherent oscillations become weaker, finally disappearing at the temperature of 79 K in the time-domain. This observation is out of expectation being in striking contrast to the case of n-type GaAs.

In order to quantitatively evaluate the transient reflectivity changes, we fit them to two damped harmonic oscillations

$$\left(\frac{\Delta R_{eo}}{R_0}\right) = \sum_i A_i exp\left(-\frac{t}{\tau_i}\right) sin(\omega_i t + \varphi_i) \qquad (1)$$

where $\tau_i$ is the decay time, $\omega_i$ is the mode frequency, and $A_i$ and $\varphi_i$ are the oscillatory amplitude and the initial phase at zero delay, respectively, as shown in Fig. 1(a) and Fig. 1(b) (blue solid lines). FFT of the fitted curves reproduces perfectly experimental results in Fig. 1(c) and Fig. 1(d). Phonon parameters obtained by the fitting versus temperature are displayed in Fig. 2. It is clear that for n-type GaAs only the decay time of LO phonons is strongly temperature dependent due to lattice anharmonicity, as mentioned above (see Fig. 2(a)). The decay rate, $1/\tau$, can be well fitted (blue dashed lines) by the Klemens channel

$$\frac{1}{\tau} = \Gamma_0(1 + 2n_{LA}) \qquad (2)$$

in which optical phonons decay into two acoustic phonons of $\omega_{LA} = \omega_{LO}/2$ with equal but opposite wave vectors [29]. Here $\Gamma_0$ is the fitting parameter, $n_{LA} = [exp(\hbar\omega_{LA}/2k_BT) - 1]^{-1}$ is the Bose-Einstein factor, and $k_B$ is the Boltzmann constant. The fitted $\Gamma_0$ is 0.26 ps$^{-1}$. The LO phonon frequency slightly decreases with the raised temperature, resulting from the thermal expansion. The frequency and decay time of LOPC modes are more or less temperature-independent in this temperature range. It might be because of the plasmon-like character of the coupled mode, in which the plasmon is less sensitive to the temperature. For p-type GaAs, however, both LO phonons and LOPC modes depend notably on the sample temperature. As shown in Fig. 2(b), the amplitude of both modes decreases monotonically to zero by cooling. This is distinct from the case of n-type GaAs in which the oscillatory amplitude is almost independent of temperature. The frequency of LOPC mode, on the other hand, increases with the decreasing temperature as well as that of LO phonons. The magnitude of the frequency shift, however, is much larger for LOPC mode, which cannot be accounted by thermal effect. The decay time of these two modes shows similar tendency as that of n-type GaAs but with a larger fluctuation.

To understand the experimental results, it is instructive to recall the generation mechanism of coherent oscillations in polar semiconductors, which has been well accepted as ultrafast screening of surface field [20]. In the polar semiconductor, the conduction and valence bands are bent at the surface due to Fermi-level pinning at charged surface states, resulting in a macroscopic surface field perpendicular to the sample surface. When electron-hole pairs are created by pumping in the depletion layer, they drift in opposite directions to compensate the intrinsic surface field. As a result, the lattice undergoes a sudden change of constraining force and starts to oscillate around a new equilibrium position, as illustrated in Fig. 3. As a result, the driving force for the excitation of coherent oscillations is

$$F(t) = R_{jkl}E_kE_l - \frac{e^*}{\varepsilon_\infty\varepsilon_0}\left[\chi^{(2)}_{jkl}E_kE_l + \chi^{(3)}_{jklm}E_kE_lE_m + \int_{-\infty}^{t}J(t')dt'\right] \quad (3)$$

where $e^*$ is the effective lattice charge, $\varepsilon_\infty$ is the high-frequency dielectric constant, $\varepsilon_0$ is the vacuum permittivity, $E_{j,k,l,m}$ is the pump filed component, the subscripts $j,k,l$ denote the Cartesian coordinates, $R_{jkl}$ is the Raman tensor, $\chi^{(2)}_{jkl}$ and $\chi^{(3)}_{jklm}$ are the second- and third-order nonlinear susceptibilities, and $J(t)$ is a time-varying current associated with the drift of photoexcited carriers in the depletion layer.

It has been experimentally shown that the oscillatory amplitude of doped GaAs is independent on the pump polarization [9]. Therefore, the anisotropic contribution from the first three terms in Eq. (3) is negligible. The driving force originates mainly from the carrier drift induced current in the depletion layer, the duration of which is determined by the diffusion time of the minority photoexcited carriers due to the photo-Dember effect. Recent experiments have unambiguously demonstrated that the oscillatory amplitude depends on the duration of driving force [30]. Coherent oscillations can only be excited when the force duration is shorter than the half period of phonon mode. Using the Schottky barrier model, we can estimate the depletion layer depth of a doped semiconductor

$$h_{dep} = \sqrt{\frac{2\varepsilon_0 \varepsilon_s V_{bi}}{qN_{n,p}}} \qquad (4)$$

where $\varepsilon_s$ is the static dielectric constant, $q$ is the electron charge, $V_{bi}$ is the built-in potential [20]. The obtained depth is 32 nm and 12 nm for n- and p-type GaAs studied here, respectively. The carrier diffusion time in the depletion layer is thus estimated by

$$\tau = \frac{h_{dep}^2}{4D_{e,h}} \qquad (5)$$

At room temperature, the diffusion coefficients $D$ for electrons and holes are 200 cm$^2$/s and 10 cm$^2$/s [31], respectively, leading to the diffusion time of 13 fs and 36 fs for n- and p-type GaAs.

Hence, the screening time of the intrinsic surface field in p-type GaAs is close to the half period (58 fs) of LO phonons. It has been well known that the depth of the depletion layer increases with the decreased temperature [32]. We thus expect that cooling may make the screening time longer, being comparable to the half period of LO phonons, by only slightly increasing the depth of the depletion layer in p-type GaAs to 15 nm, such that coherent oscillations of LO phonons cannot be excited any more. Consequently, the coupling of coherent LO phonons and plasmons cannot occur due to the delayed formation of coupled modes. For n-type GaAs, however, breaking the conditions for the generation of coherent LO phonons require more than doubling the depletion layer, which seems impossible in the investigated temperature range. A slight modulation of the depletion layer induces only a negligible change of the screening time, and as a result there is almost no change in the coherent amplitude as shown in Fig. 2(a). Here, we assume the diffusion coefficients of photoexcited carriers to be constant at the various lattice temperature. This assumption works at the very early time of the laser-matter interaction during which the photoexcited carriers are not thermally equilibrated and the lattices do not involve into the photoexcited carrier diffusion.

In order to qualitatively support this argument, we assume the temperature dependence of the depletion layer thickness $h_{dep}(T) = h_{dep}^0 - \alpha(T - T_0)$ as the first-order approximation for both n- and p-type GaAs, where $h_{dep}^0$ is the thickness at room temperature $T_0$ and $\alpha$ is the fitting parameter. The carrier diffusion times at different temperature, $\tau(T)$, therefore, can be estimated by Eq. (5). Then we solved the equation of motion for a damped harmonic oscillator at the LO frequency that is driven by a current force described in Eq. (3) at different temperature. Here, $J(t)$ takes the form of a Gaussian function characterized by the rise time, $\tau_r(T)$, that is proportional to the carrier diffusion time $\tau(T)$, thus leading to the driving force $F(t)$ of the form of an error function. With the fitting parameter $\alpha$ of 0.033 nm/K, the calculated oscillatory amplitude as a function of temperature is shown in Fig. 4 for both n- and p-type GaAs, in which the amplitude is normalized by the unity at room temperature. As expected, the calculations reproduce well the experimental results for both samples, though the model employed is rather simple. Indeed, due to a lower mobility of holes the depletion layer thickness and, as a result, the screening time is very close to the upper limit of coherent phonon generation, whereas for electrons due to their high mobility the screening time is significantly shorter than the upper limit. Therefore, the temperature changes strongly affect the coherent phonon generation in p-type sample, leaving the generation conditions in n-type sample almost unchanged.

The frequency of LOPC modes in p-type GaAs is modulated by the lattice temperature. We attribute this modulation to the temperature dependence of the plasmon damping constant. In the critical damping condition which is suitable to the case of p-type GaAs, the frequency of coupled modes strongly depends on the damping constant of the plasmon mode [8].

**IV. Summary**

To summarize, the effect of temperature on the coupling between coherent LO phonons and plasmons has been investigated in both n- and p-type GaAs by using ultrafast coherent phonon spectroscopy. A distinctly different dependence of mode properties on the temperature has been observed in two samples, in which p-type GaAs exhibits anomalous behavior, being strikingly different from n-type GaAs and other materials. This anomaly could be in principle understood by the increased screening time of surface field in p-type GaAs at lower temperature resulting in a suppression of the generation of coherent oscillations. In this case, the temperature dependence observed here suggests that ultrafast optical phonon spectroscopy may be used to characterize the depth of the depletion layer in a non-contact and high-speed way.

**Acknowledgment**

The authors thank Dr. Katsura Norimatsu and Arihiro Goto for experimental assistance. This work was supported in part by China 1000-Young Talents Plan. OVM acknowledges the support from the Russian Foundation for Basic Research (No. 17-02-00002) and the Government of the Russian Federation (No. 05.Y09.21.0018). KGN acknowledges the support from Japan Society for the Promotion of Science, Grants-in-Aid for Scientific Research (No. 17K19051, 17H02797, 15H02103, and 15K13377).

**Figure captions**

**Fig. 1** Transient reflectivity changes (red solid lines) induced by the ultrafast excitation at different temperatures for n-type (a) and p-type GaAs (b). Blue solid lines in (a) and (b) are fitted curves by Eq. (1). FFTs of the time-domain signals are shown in (c) and (d) for n- and p-type GaAs, respectively, where symbols are experimental results and solid lines are fitted results.

**Fig. 2** Temperature dependence of mode properties obtained by fitting to Eq. (1) for n-type (a) and p-type GaAs (b). Blue and red circles indicate coherent LO phonons and LOPC modes, respectively. The blue dashed lines in the panel of decay time are obtained by fitting the data of LO phonons to the Klemens channel.

**Fig. 3** Transient screening of surface field for generating coherent oscillations of LO phonons in n-type (a) and p-type GaAs (b). Note that the depth of the depletion layer and the diffusion coefficient of photoexcited carriers are different for two samples. $t_0$ and $t_1$ are the moments before and after pumping. The arrow size indicates qualitatively the drift velocity of photoexcited carriers.

**Fig. 4** Calculated (dashed lines) and measured (circles) oscillatory amplitude of coherent LO phonons as a function of temperature for both n-type (red) and p-type GaAs (blue). The amplitude has been normalized by the one at room temperature. Details of the calculation are described in the main text.

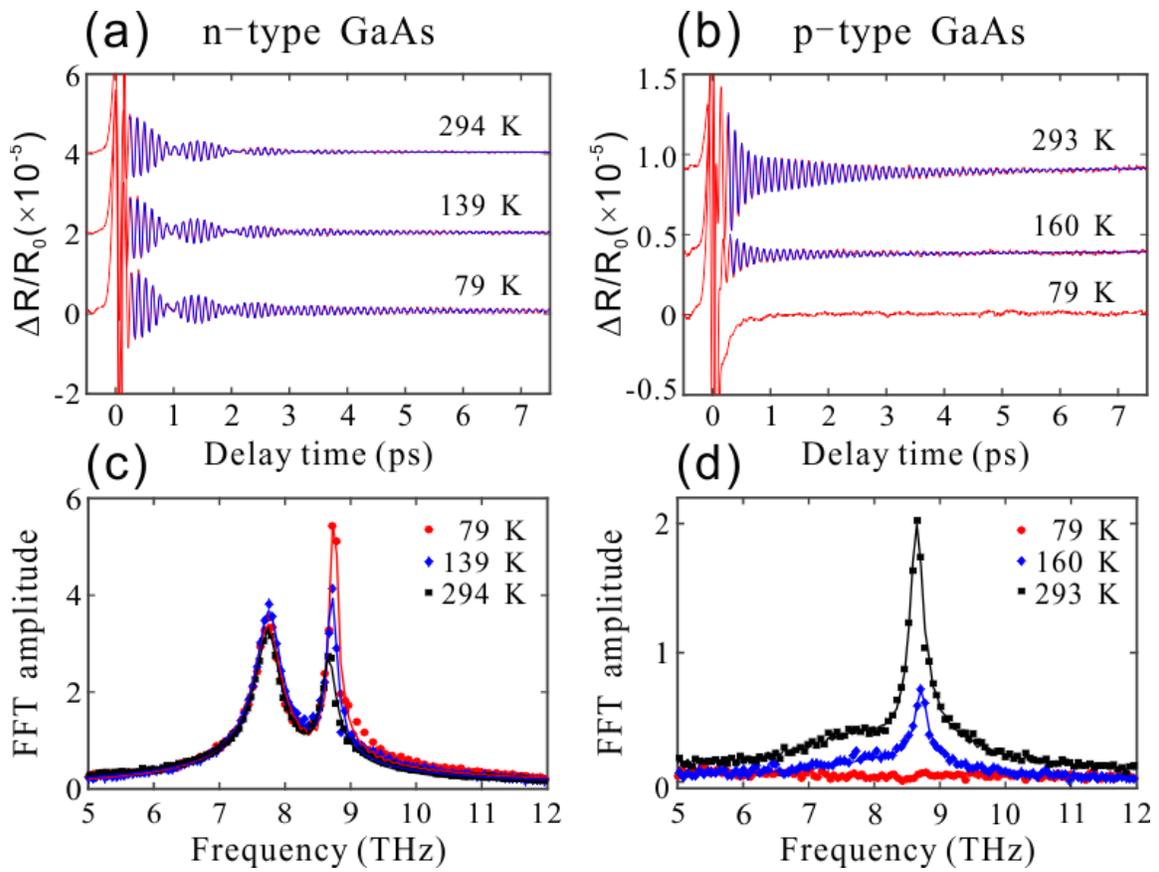

**Fig. 1** Hu *et al*.

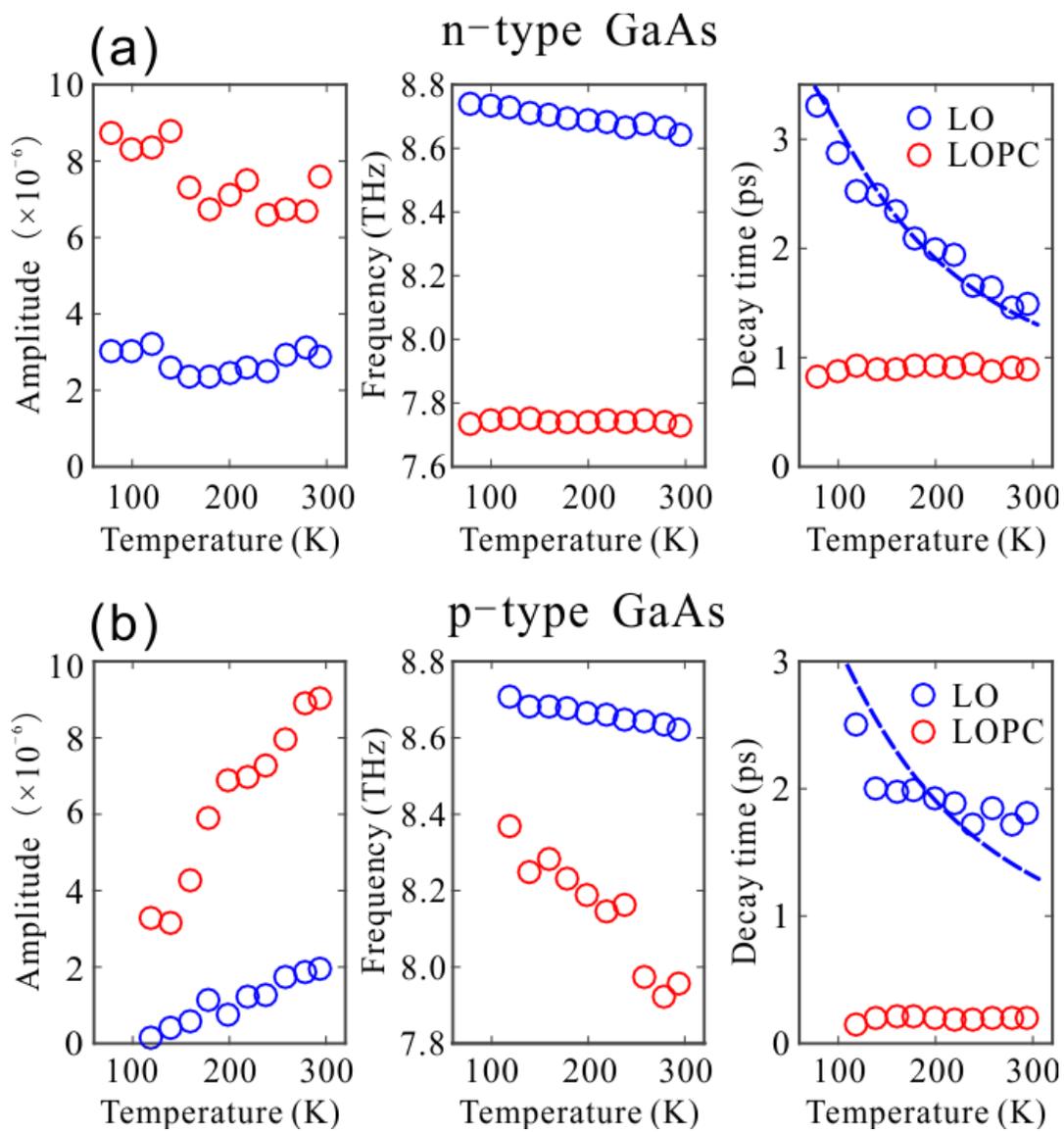

**Fig. 2** Hu *et al*.

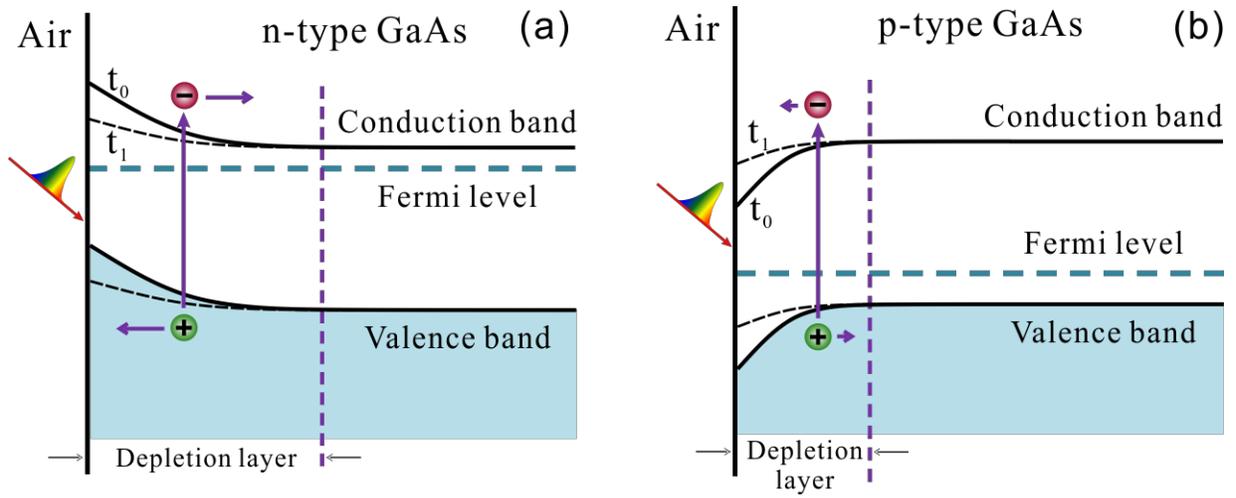

**Fig. 3** Hu *et al*.

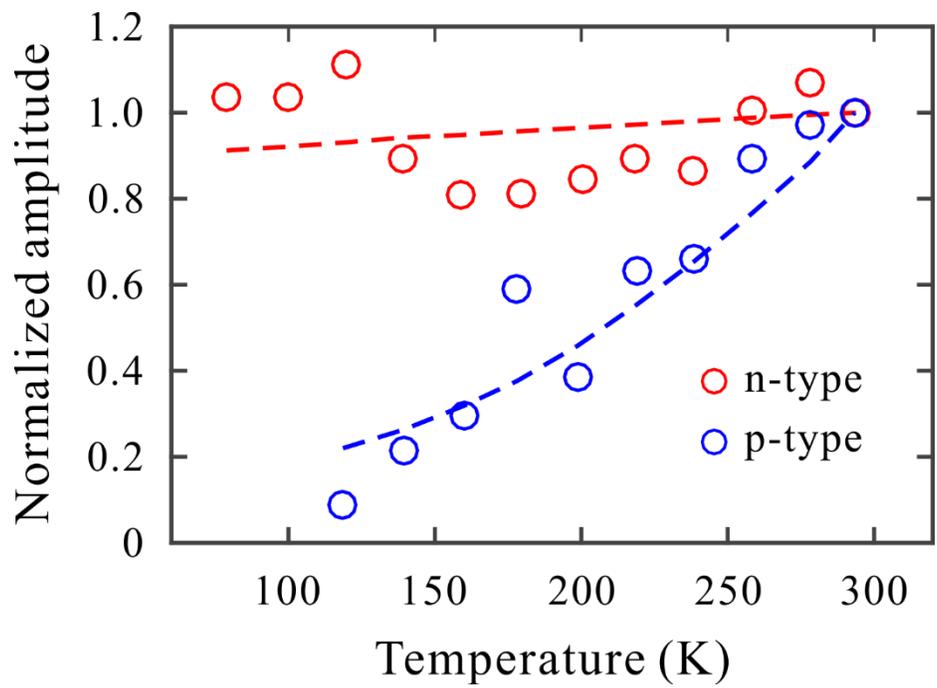

**Fig. 4** Hu *et al.*